\documentclass{article} 
\usepackage{amsfonts,amssymb}
\usepackage{stmaryrd}
\usepackage{amsmath}
\usepackage{color}
\usepackage{xy}
\xyoption{all}
\CompileMatrices

\newdimen\proofrulebreadth \proofrulebreadth=.05em
\newdimen\proofdotseparation \proofdotseparation=1.25ex
\newdimen\proofrulebaseline \proofrulebaseline=2ex
\newcount\proofdotnumber \proofdotnumber=3
\let\then\relax
\def\hfi{\hskip0pt plus.0001fil}
\mathchardef\squigto="3A3B
\newif\ifinsideprooftree\insideprooftreefalse
\newif\ifonleftofproofrule\onleftofproofrulefalse
\newif\ifproofdots\proofdotsfalse
\newif\ifdoubleproof\doubleprooffalse
\let\wereinproofbit\relax
\newdimen\shortenproofleft
\newdimen\shortenproofright
\newdimen\proofbelowshift
\newbox\proofabove
\newbox\proofbelow
\newbox\proofrulename
\def\shiftproofbelow{\let\next\relax\afterassignment\setshiftproofbelow\dimen0 }
\def\shiftproofbelowneg{\def\next{\multiply\dimen0 by-1 }%
\afterassignment\setshiftproofbelow\dimen0 }
\def\setshiftproofbelow{\next\proofbelowshift=\dimen0 }
\def\setproofrulebreadth{\proofrulebreadth}

\def\prooftree{
\ifnum  \lastpenalty=1
\then   \unpenalty
\else   \onleftofproofrulefalse
\fi
\ifonleftofproofrule
\else   \ifinsideprooftree
        \then   \hskip.5em plus1fil
        \fi
\fi
\bgroup
\setbox\proofbelow=\hbox{}\setbox\proofrulename=\hbox{}%
\let\justifies\proofover\let\leadsto\proofoverdots\let\Justifies\proofoverdbl
\let\using\proofusing\let\[\prooftree
\ifinsideprooftree\let\]\endprooftree\fi
\proofdotsfalse\doubleprooffalse
\let\thickness\setproofrulebreadth
\let\shiftright\shiftproofbelow \let\shift\shiftproofbelow
\let\shiftleft\shiftproofbelowneg
\let\ifwasinsideprooftree\ifinsideprooftree
\insideprooftreetrue
\setbox\proofabove=\hbox\bgroup$\displaystyle 
\let\wereinproofbit\prooftree
\shortenproofleft=0pt \shortenproofright=0pt \proofbelowshift=0pt
\onleftofproofruletrue\penalty1
}

\def\eproofbit{
\ifx    \wereinproofbit\prooftree
\then   \ifcase \lastpenalty
        \then   \shortenproofright=0pt  
        \or     \unpenalty\hfil         
        \or     \unpenalty\unskip       
        \else   \shortenproofright=0pt  
        \fi
\fi
\global\dimen0=\shortenproofleft
\global\dimen1=\shortenproofright
\global\dimen2=\proofrulebreadth
\global\dimen3=\proofbelowshift
\global\dimen4=\proofdotseparation
\global\count255=\proofdotnumber
$\egroup  
\shortenproofleft=\dimen0
\shortenproofright=\dimen1
\proofrulebreadth=\dimen2
\proofbelowshift=\dimen3
\proofdotseparation=\dimen4
\proofdotnumber=\count255
}

\def\proofover{
\eproofbit 
\setbox\proofbelow=\hbox\bgroup 
\let\wereinproofbit\proofover
$\displaystyle
}%
\def\proofoverdbl{
\eproofbit 
\doubleprooftrue
\setbox\proofbelow=\hbox\bgroup 
\let\wereinproofbit\proofoverdbl
$\displaystyle
}%
\def\proofoverdots{
\eproofbit 
\proofdotstrue
\setbox\proofbelow=\hbox\bgroup 
\let\wereinproofbit\proofoverdots
$\displaystyle
}%
\def\proofusing{
\eproofbit 
\setbox\proofrulename=\hbox\bgroup 
\let\wereinproofbit\proofusing
\kern0.3em$
}

\def\endprooftree{
\eproofbit 
  \dimen5 =0pt
\dimen0=\wd\proofabove \advance\dimen0-\shortenproofleft
\advance\dimen0-\shortenproofright
\dimen1=.5\dimen0 \advance\dimen1-.5\wd\proofbelow
\dimen4=\dimen1
\advance\dimen1\proofbelowshift \advance\dimen4-\proofbelowshift
\ifdim  \dimen1<0pt
\then   \advance\shortenproofleft\dimen1
        \advance\dimen0-\dimen1
        \dimen1=0pt
        \ifdim  \shortenproofleft<0pt
        \then   \setbox\proofabove=\hbox{%
                        \kern-\shortenproofleft\unhbox\proofabove}%
                \shortenproofleft=0pt
        \fi
\fi
\ifdim  \dimen4<0pt
\then   \advance\shortenproofright\dimen4
        \advance\dimen0-\dimen4
        \dimen4=0pt
\fi
\ifdim  \shortenproofright<\wd\proofrulename
\then   \shortenproofright=\wd\proofrulename
\fi
\dimen2=\shortenproofleft \advance\dimen2 by\dimen1
\dimen3=\shortenproofright\advance\dimen3 by\dimen4
\ifproofdots
\then
        \dimen6=\shortenproofleft \advance\dimen6 .5\dimen0
        \setbox1=\vbox to\proofdotseparation{\vss\hbox{$\cdot$}\vss}%
        \setbox0=\hbox{%
                \advance\dimen6-.5\wd1
                \kern\dimen6
                $\vcenter to\proofdotnumber\proofdotseparation
                        {\leaders\box1\vfill}$%
                \unhbox\proofrulename}%
\else   \dimen6=\fontdimen22\the\textfont2 
        \dimen7=\dimen6
        \advance\dimen6by.5\proofrulebreadth
        \advance\dimen7by-.5\proofrulebreadth
        \setbox0=\hbox{%
                \kern\shortenproofleft
                \ifdoubleproof
                \then   \hbox to\dimen0{%
                        $\mathsurround0pt\mathord=\mkern-6mu%
                        \cleaders\hbox{$\mkern-2mu=\mkern-2mu$}\hfill
                        \mkern-6mu\mathord=$}%
                \else   \vrule height\dimen6 depth-\dimen7 width\dimen0
                \fi
                \unhbox\proofrulename}%
        \ht0=\dimen6 \dp0=-\dimen7
\fi
\let\doll\relax
\ifwasinsideprooftree
\then   \let\VBOX\vbox
\else   \ifmmode\else$\let\doll=$\fi
        \let\VBOX\vcenter
\fi
\VBOX   {\baselineskip\proofrulebaseline \lineskip.2ex
        \expandafter\lineskiplimit\ifproofdots0ex\else-0.6ex\fi
        \hbox   spread\dimen5   {\hfi\unhbox\proofabove\hfi}%
        \hbox{\box0}%
        \hbox   {\kern\dimen2 \box\proofbelow}}\doll%
\global\dimen2=\dimen2
\global\dimen3=\dimen3
\egroup 
\ifonleftofproofrule
\then   \shortenproofleft=\dimen2
\fi
\shortenproofright=\dimen3
\onleftofproofrulefalse
\ifinsideprooftree
\then   \hskip.5em plus 1fil \penalty2
\fi
}

\newcounter{countroman}
\newenvironment{rnumerate}%
{\begin{list}{{\rm (\roman{countroman})}}{\usecounter{countroman}}}%
{\end{list}}

\newcounter{countalpha}
{\begin{list}{(\alph{countalpha})}{\usecounter{countalpha}}}%
{\end{list}}

\newcounter{countalphabf}
{\protect\begin{list}{{\rm (}{\bf \protect\alph{countalphabf}}{\rm%
)}}{\protect\usecounter{countalphabf}}}%
{\end{list}}

\mathcode`\<="4268 
\mathcode`\>="5269 
\mathchardef\gt="313E 
\mathchardef\lt="313C 

\newcommand{\id}{{\rm id}}

\newcommand{\FHilb}{{\sf FHilb}}

\newcommand{\Rel}{{\sf Rel}}

\newcommand{\Set}{{\sf Set}}

\newcommand{\Span}{{\sf Span}}
\newcommand{\Pred}{{\sf P}}
\newcommand{\Prop}{{\PPP}}
\newcommand{\Prob}{{\PPp}}
\newcommand{\RTst}{{\TTT}}
\newcommand{\HTst}{{\sf T}}
\newcommand{\MTst}{{\TTt}}

\newcommand{\parr}{\bindnasrepma}

\renewcommand{\to}{\xymatrix@C-.5pc{\ar[r]&}}
\newcommand{\ot}{\xymatrix@C-.5pc{& \ar[l]}}
\newcommand{\tto}[1]{\xymatrix@C-.5pc{\ar[r]^{#1}&}}
\newcommand{\oot}[1]{\xymatrix@C-.5pc{&\ar[l]_{#1}}}
\newcommand{\mono}{\xymatrix@C-.5pc{\ar@{>->}[r]&}} 
\newcommand{\epi}{\xymatrix@C-.5pc{\ar@{->>}[r]&}}
\newcommand{\mmono}[1]{\xymatrix@C-.5pc{\ar@{>->}[r]^{#1}&}} 
\newcommand{\eepi}[1]{\xymatrix@C-.5pc{\ar@{->>}[r]^{#1}&}}
\renewcommand{\mapsto}{\xymatrix@C-.5pc{\ar@{|->}[r]&}}
\newcommand{\mmapsto}[1]{\xymatrix@C-.5pc{\ar@{|->}[r]^{#1}&}}
\newcommand{\inclusion}{\xymatrix@C-.5pc{\ar@{^{(}->}[r] &}}
\newcommand{\iinclusion}[1]{\xymatrix@C-.5pc{\ar@{^{(}->}[r]^{#1}&}}
\newcommand{\rrel}[1]{\xymatrix@C-.5pc{\ar[r]|{|}^{#1}&}}
\newcommand{\rel}{\xymatrix@C-.5pc{\ar[r]|{|}&}}
\newcommand{\ler}{\xymatrix@C-.5pc{&\ar[l]|{|}}}
\newcommand{\ller}[1]{\xymatrix@C-.5pc{&\ar[l]|{|}_{#1}}}

\newcommand{\AAA}{{\cal A}}
\newcommand{\BBB}{{\cal B}}
\newcommand{\CCC}{{\cal C}}

\newcommand{\EEE}{{\cal E}}

\newcommand{\HHH}{{\cal H}}

\newcommand{\KKK}{{\cal K}}

\newcommand{{\MMM}}{{\cal M}}

\newcommand{{\PPP}}{{\cal P}}

\newcommand{\RRR}{{\cal R}}

\newcommand{\TTT}{{\cal T}}

\renewcommand{\Bbb}{\mathbb}

\newcommand{\CCc}{{\Bbb C}}

\newcommand{\NNn}{{\Bbb N}}

\newcommand{\PPp}{{\Bbb P}}

\newcommand{\TTt}{{\Bbb T}}

\newcommand{\ZZz}{{\Bbb Z}}

\newcommand{\WP}{\mbox{\Large $\wp$}}

 \def\pushright#1{{
    \parfillskip=0pt            
    \widowpenalty=10000         
    \displaywidowpenalty=10000  
    \finalhyphendemerits=0      
    \leavevmode                 
    \unskip                     
    \nobreak                    
    \hfil                       
    \penalty50                  
    \hskip.2em                  
    \null                       
    \hfill                      
    {#1}                        
    \par}}                      

 \def\qed{\pushright{$\square$}\penalty-700 \smallskip}

\newenvironment{prf}[1]{\begin{trivlist} \item[{\bf ~Proof}#1.]}%
{\qed\end{trivlist}}

\newcommand{\be}[1]{\begin{#1}}

\newcommand{\ee}[1]{\end{#1}}
\newcommand{\beq}{\begin{equation}}
\newcommand{\eeq}{\end{equation}}
\newcommand{\ba}[1]{\begin{array}{#1}}
\newcommand{\ea}{\end{array}}
\newcommand{\bea}{\begin{eqnarray}}
\newcommand{\eea}{\end{eqnarray}}
\newcommand{\bear}{\begin{eqnarray*}}
\newcommand{\eear}{\end{eqnarray*}}
\newcommand{\bpr}{\begin{prf}{}}
\newcommand{\epr}{\end{prf}}
\newcommand{\bprf}[1]{\begin{prf}{#1}}
\newcommand{\eprf}{\end{prf}}
\newcommand{\midd}[1]{\{#1\}}
\newcommand{\rays}[1]{\widehat{#1}}

\newtheorem{thm}{Theorem}[section]
\newtheorem{defn}[thm]{Definition}

\newtheorem{cond}{}[thm]

\newcommand{\from}{\!\in\!}
\newcommand{\DA}{dagger mix autonomous}
\newcommand{\DDA}{Dagger mix autonomous}

\begin{document}

  \title{Relating toy models of quantum computation:\\
  comprehension, complementarity\\ and \DA\  categories} 
  \author{Dusko Pavlovic \thanks{Supported by ONR.}\\  
  Kestrel Institute and Oxford University\\
  {\tt dusko@\{kestrel.edu,comlab.ox.ac.uk\}}} 
  
   \date{}

  \maketitle

\begin{abstract} 
Toy models have been used to separate important features of quantum computation from the rich background of the standard Hilbert space model. Category theory, on the other hand, is a general tool to separate components of mathematical structures, and analyze one layer at a time. It seems natural to combine the two approaches, and several authors have already pursued this idea. We explore {\em categorical comprehension construction\/} as a tool for adding features to toy models. We use it to comprehend quantum propositions and probabilities within the basic model of finite-dimensional Hilbert spaces. We also analyze complementary quantum observables over the category of sets and relations. This leads into the realm of {\em test spaces}, a well-studied model. We present one of many possible extensions of this model, enabled by the comprehension construction. Conspicuously, all models obtained in this way carry the same categorical structure, {\em extending\/} the familiar dagger compact framework with the complementation operations. We call the obtained structure {\em \DA\ }, because it extends mix autonomous categories, popular in computer science, in a similar way like dagger compact structure extends compact categories. \DDA\  categories seem to arise quite naturally in quantum computation, as soon as complementarity is viewed as a part of the global structure. 
\end{abstract}


\section{Introduction, background, related work}
Mathematical models of physical systems are often complicated. Quantum physics in particular is built over very rich mathematical structures. The efforts to extract conceptual components from these structures, and to analyze the particular quantum phenomena supported in such fragments, can be traced back all the way to Birkhoff and von Neumann. Nowadays, such efforts sometimes lead to {\em toy models} \cite{Foulis-Randall:test,Mermin:moon,SpekkensR:toy,Coecke-Edwards,AbramskyS:BigToy}.

But components are only useful if they can be used to build something. Isolating some quantum phenomena in partial models is only useful if we know how to combine these partial models together, in order to relate the analyzed phenomena; and how to incrementally build larger pictures  of quantum theory from smaller fragments. Categorical tools seem well suited for this purpose. Besides providing a categorical view of quantum programming in the standard Hilbert space model \cite{SelingerP:QPL,SelingerP:HOQC}, categorical semantics of quantum computation \cite{Abramsky-Coecke:LICS,PavlovicD:QMWS,Coecke-Duncan,PavlovicD:CQStruct} can be viewed as a toolkit for building, combining and reconstructing toy models of quantum computation, and nonstandard models in general. In this spirit, Coecke, Edwards have reconstructed and extended Spekkens' toy model in a categorical framework \cite{Coecke-Edwards,Coecke-Edwards-Spekkens}. Abramsky used the Chu construction as a categorical tool for building {\em big\/} toy models, encompassing not only quantum computation, but possibly other exotic kinds of systems \cite{AbramskyS:BigToy,AbramskyS:ToyChu}. Our exploration here can be viewed as an attempt in the same direction: we propose another categorical construction that might be useful as a piece of the toolkit of categorical semantics of quantum computation.

The starting point of the path towards this quantum categorical toolkit was the remarkably simple observation, due to Abramsky and Coecke \cite{Abramsky-Coecke:LICS}, that a basic form of quantum entanglement can be modeled using the duality structure of {\em compact categories\/} \cite{Kelly-Laplaza}. Extended with an additional operation, the contravariant functor {\em dagger $\ddag$}, corresponding to the operator adjunction, {\em dagger compact categories\/} were thus proposed as a basic type system for an abstract view of quantum computation. The abstract characterizations of mixed quantum states as completely positive operators \cite{SelingerP:CP}, and of quantum and classical observables in terms of special Frobenius algebras \cite{PavlovicD:QMWS,PavlovicD:MSCS08}, as well as some related algebraic structures \cite{Coecke-Duncan,Coecke-Kissinger,PavlovicD:Qabs}, were soon added to the quantum categorical toolkit, allowing simple characterizations of many quantum operations \cite{PavlovicD:CQStruct}. 

In this note, we consider a categorical tool for incremental refinement of toy models, which thus allows by adding new features, such as quantum propositions, probabilities, or complementarity. In the Hilbert space model, quantum propositions are represented as closed subspaces. Quantum logic, initiated by Birkhoff and von Neumann \cite{Birkhoff-vonNeumann:LQM}, was an attempt to capture the logical content of quantum theory by axiomitizing such propositions, and studying them algebraically. The resulting lattice theory captures  some important aspects of quantum theory, but abstracts away some other important aspects. Nevertheless, the link with quantum probability theory through Gleason's theorem \cite{Gleason} is undoubtedly of great conceptual importance.

So how can we add quantum propositions to a toy model, viewed as a dagger compact category? The categorical construction that can be used generalizes the familiar set theoretic schema of {\em comprehension}. It is briefly described in Sec.~\ref{Comprehension}. In the rest of the paper, we apply this {\em categorical comprehension construction\/} to simple examples, and build categories of quantum systems where the quantum operations are required to preserve the comprehended structure: e.g., quantum propositions, observables, etc. The operations that come with these added structures, echoing Birkhoff-von Neumann's logics, are reflected in the structure of the obtained categories. In Sec.~\ref{propositions}, we finally come around to the original task of adding quantum propositions, and adjoin quantum propositions as explicit structure to finitely dimensional Hilbert spaces. In Sec.~\ref{Autonomous}, we describe the resulting categorical structure and define {\em \DA\  categories}. In Sec.~\ref{Complementarity}, we discuss the ways to adjoin complementary observables to two basic models, again using the comprehension construction. In Sec.~\ref{More} we describe, very briefly, two slightly richer toy models arising from the same construction, just to give an idea of the possibilities that it opens. Sec.~\ref{Future} lists some open questions that arise from it.

\section{Categorical comprehension construction}\label{Comprehension}
The set theoretic comprehension principle asserts that predicates over a set $S$ are in one-to-one correspondence with the subsets of $S$:
\[
\prooftree
\Phi \ :\ S\to 2
\justifies
\{x\from S\ |\  \Phi(x)\} \hookrightarrow S
\endprooftree
\]
The topological generalization of this correspondence establishes the equivalence between \'etal\'e spaces over a space $S$ and sheaves of sets under $S$ \cite{MacLane-Moerdijk}. The categorical generalizations go back to {\em Grothendieck's construction\/} of the discrete fibration corresponding to a sheaf [{\em ibidem}]; but their logical interpretation, and the connection with the idea of comprehension is due to Lawvere \cite{LawvereFW:comprehension,PavlovicD:thesis,JacobsB:book}. 

In the most general form, originally outlined in \cite{PavlovicD:SIC}, the categorical comprehension schema establishes the correspondence of lax functors from a category $\CCC$ to the bicategory $\Span$ and arbitrary (small-fibered) functors to $\CCC$
\[
\prooftree
{\Pred} \ :\ \CCC \to \Span
\justifies
\int_{\CCC} {\Pred} \to \CCC
\endprooftree
\]
To explain this correspondence, we first describe the bicategory $\Span$ and the notion of lax functors to it, and then specify the comprehension construction $\int_{\CCC} {\Pred}$.

\begin{defn}
The bicategory 
$\Span$ consists of
\begin{itemize}
\item sets $A,B,\ldots$ as objects (0-sets);
\item a morphism (1-cell) $F:A\to B$ is a span of functions $A\ot F\to B$;
\item a transformation (2-cell) $\chi: F\to G:A\to B$ is a function $F\tto{\chi} G$ such that both triangles in the following diagram commute.
\beq\label{transfromation}
\xymatrix@-1pc{
	&&\ar[dll] F\ar[drr] \ar[dd]^\chi \\
A  &&&& B\\
&& \ar[ull] G \ar[urr]
}
\eeq
\end{itemize}
While the composition of transformations is obvious, the composition of spans is induced by pullbacks:
\beq\label{spancomp}
\xymatrix@-1pc{&&&& \ar[dll] (F;H) \ar[drr]\\
	&&\ar[dll] F\ar[drr] &&&& H \ar[dll] \ar[drr]\\
A  &&&& B &&&& C}
\eeq
\end{defn}

\noindent{\bf Remark.} If a span $A\oot{\pi_A} F\tto{\pi_B} B$ is viewed  as a set matrix $F \in \Set^{A\times B}$, with the entries $F_{ab} = <\pi_A,\pi_B>^{-1}(a,b)$, then the span composition becomes the usual matrix composition.

\begin{defn}
A\/ {\em (comprehension) specification\/} is a lax functor  ${\Pred} :\CCC\to\Span$, consisting of the following assignements: 
\begin{itemize}
\item for each object $A\in |\CCC|$ a set ${\Pred} A$,
\item for each morphism $A\tto{f} B$ a span ${\Pred} A\ot \midd f\to {\Pred} B$, and moreover
\item for every composable pair $A\tto{f} B\tto{g} C$ a transformation $\mu_{fg}$ 
\beq\label{mu}
\xymatrix@-1pc{ &&&& \midd{f\, ; g} \ar@/_/[ddddllll] \ar@/^/[ddddrrrr] \\ \\
			&&&& \ar[dll] \midd{f}\, ; \midd{g} \ar[drr] \ar[uu]_{\mu_{fg}}\\
	&&\ar[dll] \midd f \ar[drr] &&&& \midd g \ar[dll] \ar[drr]\\
{\Pred} A  &&&& {\Pred} B &&&& {\Pred} C
}
\eeq
\item for every object $A\in |\CCC|$ a transformation $\eta_A$  
\beq\label{eta}
\xymatrix@-1pc{
	&&\ar[dll] \midd{\id_A}\ar[drr]  \\
{\Pred} A  &&&& {\Pred} A\\
&& \id_{{\Pred} A} \ar[ull]  \ar[urr] \ar[uu]^{\eta_A}
}
\eeq
such that the following diagrams commute
\[
\xymatrix@C-1.8pc@R-.8pc{
\midd f ; \midd g ; \midd h \ar[rrrrrrrr]^{\mu_{fg};\midd h} \ar[dddd]_{\midd f;\mu_{gh}}&&&&&&&& \midd{f;g};\midd h \ar[dddd]^\mu 
\\ \\
\\ \\
\midd{f}; \midd{g;h} \ar[rrrrrrrr]_{\mu} &&&&&&&& \midd{f;g;h} 
\\ \\
\midd f \ar@{=}[r] \ar@{=}[d] \ar[rrrrrrrrdddd]^{\id}& \midd{f};\id_{{\Pred} B} \ar[rrrrrrr]^{f; \eta_B} &&&&&&& \midd f ; \midd{\id_B} \ar[dddd]^\mu \\
\id_{{\Pred} A};\midd f \ar[ddd]_{\eta_A;\midd {f}} 
\\ \\ \\
\midd{\id_A};\midd{f} \ar[rrrrrrrr]_\mu &&&&&&&& \midd{f}
}
\]
\end{itemize}
\end{defn}

\noindent{\bf Remark.} When spans are viewed as matrices of sets, then the above data become the families:
\bea
\sum_{\beta\in {\Pred} B} 
\alpha \midd f \beta \times \beta \midd g \gamma & 
\xymatrix{ \ar[rr]^{\mu_{fg}^{\alpha\gamma}}&& \hspace{1ex}}
& \alpha \midd{f;g} \gamma \label{moo}\\
1 & \xymatrix{ \ar[rr]^{\eta_A^\alpha}&& \hspace{1ex}}
& \alpha \midd{\id_A} \alpha \label{eeta}
\eea
indexed by $\alpha\in {\Pred} A$ and $\gamma \in {\Pred} C$, and where e.g. $\alpha\midd f \beta = <\pi_{{\Pred} A},\pi_{{\Pred} B}>^{-1}(\alpha,\beta)$ denotes the entry of the set matrix $\midd f \in \Set^{{\Pred} A \times {\Pred} B}$.

\begin{defn}\label{comprehendef}
The {\em comprehension\/} of a specification ${\Pred} : \CCC\to \Span $ is the functor  $\int_{\CCC} {\Pred} \to \CCC$ where the {\em comprehension category\/} $\int_\CCC {\Pred}$ is defined as follows:
\bear
\left| \int_\CCC {\Pred} \right| & = & \sum_{A\in |\CCC|} {\Pred} A\\
\int_\RRR {\Pred} \big(<A,\alpha>, <B,\beta>\big) & = & \sum_{f\in \CCC(A,B)} \alpha \midd f \beta
\eear
An arrow in $\int_\CCC {\Pred}$ is thus a pair $<f,\varphi>:<A,\alpha>\to <B,\beta>$ where $f\in \CCC(A,B)$ and $\varphi\in \alpha\midd f \beta$. The identities and the composition are:
\bear
\id_{<A,\alpha>} & = & <\id_A,\eta_A^\alpha>\\
<f,\varphi> ; <g,\psi> & = & \big<(f;g), \mu_{fg}^{\alpha\gamma}(\varphi,\psi)\big>
\eear
The comprehension functor $\int_\CCC {\Pred} \to \CCC$ is the obvious projection.
\end{defn}

\begin{defn}
A functor $F:\EEE\to \CCC$ is said to be {\em small\/} if for every object $A\in |\CCC|$ the class of objects $F^{-1}A\subseteq |\EEE|$ is a set.
\end{defn}

\noindent{\bf The correspondence.} Any small functor $E:\EEE\to \CCC$ induces a comprehension specification ${\Pred}_E:\CCC \to \Span$, defined
\bear
{\Pred} A & = & E^{-1} A\\
\alpha \midd f \beta & = & \{\varphi\in \EEE(\alpha, \beta)\ |\ E\varphi = f\} 
\eear
with $\eta_A^\alpha = \id_\alpha$ and $\mu_{fg}$ induced by the composition in $\EEE$. The obvious equivalence $\EEE \simeq \int_\CCC {\Pred}_E$ preserves the projections to $\CCC$. The functor $E:\EEE\to \CCC$ is faithful if and only if every span ${\Pred}_E A \ot \midd f \to {\Pred}_E B$ is a binary relation, i.e. $\midd f \subseteq {\Pred}_E A\times {\Pred}_E B$. Putting this all together we get the following.

\begin{thm}\label{thm}
For every category $\CCC$ there are one-to-one correspondences of 
\begin{rnumerate}
\item small functors to $\CCC$ and lax functors $\CCC\to \Span$,
\item small faithful functors to $\CCC$ and lax functors $\CCC\to \Rel$.
\end{rnumerate}
\end{thm}

\noindent{\bf Remark.} Note that composing relations as spans does not directly give relations: the assumption that the spans $F\inclusion A\times B$ and $G\inclusion B\times C$ in \eqref{spancomp} are monic does not necessarily imply that the span $(F;G)\to A\times C$ is monic. However, the image factorization $(F;G)\epi [F;G]\inclusion A\times C$ yields the relation $[F;G]$, which is the usual relational composition of the relations $F$ and $G$. If relations are viewed as matrices of 0s and 1s, this factorization replaces with 1 each nonempty set that may occur in the matrix $(F;G)$.

\section{Comprehending quantum propositions}\label{propositions}
To begin, let us comprehend quantum propositions within the finite-dimensional part of the standard model%
. The base category $\CCC$ is thus the category $\FHilb$ of finite-dimensional complex Hilbert spaces, and the specification ${\Prop} : \FHilb \to  \Rel$ simply maps each space $\HHH$ to the set ${\Prop}\HHH$ of quantum propositions, {\em viz\/} subspaces $\chi \subseteq \HHH$. Each linear map $f\in \FHilb(\HHH,\KKK)$ induces a binary relation $\midd f \subseteq {\Prop}\HHH\times {\Prop}\KKK$ such that
\bear
\chi \midd f \kappa &\iff & f \chi\subseteq \kappa
\eear
holds for $\chi \in{\Prop} \HHH$ and $\kappa \in{\Prop} \KKK$. In the posetal bicategory $\Rel$ of relations, transformations \eqref{mu} and \eqref{eta} boil down to the requirements
\bear
\chi \midd f \kappa \wedge \kappa \midd g \vartheta & \Longrightarrow & \chi \midd{f;g} \vartheta\\
\chi = \chi' & \Longrightarrow & \chi \midd{\id} \chi'
\eear
which are obviously satisfied.  The total category $\int_{\FHilb} {\Prop}$ of ${\Prop} :\FHilb \to \Rel$  consists of the pairs $<\chi\subseteq \HHH>$ as objects, and a morphism $<\chi\subseteq \HHH>\to <\kappa\subseteq \KKK>$ is simply a linear operator $f:\HHH\to\KKK$ such that $f \chi\subseteq \kappa$. 

The upshot of this construction is that the dagger compact structure of the base category $\FHilb$ and the orthomodular structure of each lattice ${\Prop} \HHH$ are now integrated in  the structure of the comprehension category $\int_\FHilb \Prop$. We spell out this structure in the next section.

\section{\DDA\  categories}\label{Autonomous}
\be{defn}
A {\em star autonomous category\/} \cite{BarrM:Staraut} is 
\begin{itemize}
\item a symmetric monoidal category $(\AAA,\otimes,\top)$ with 
\item a contravariant duality $(-)^\ast: \AAA^{op}\to \AAA$, such that
\item $\AAA(A\otimes X,\ Y^\ast) \cong \AAA\left(X,\ (A\otimes Y)^\ast \right)$, naturally in $A,X,Y$.
\end{itemize}
The induced correspondence $\AAA(X^\ast, X^\ast)\cong \AAA(X\otimes X^\ast, \top^\ast) \cong \AAA\left(X,X^{\ast\ast}\right)$ is required to map $X^\ast\tto\id {X^\ast}$ to the unit $X\tto\eta X^{\ast\ast}$ of the duality. 

The dual monoidal structure $(\AAA,\parr,\bot)$ is defined by $X\parr Y = \left(X^\ast \otimes Y^\ast\right)^\ast$, $\bot = \top^\ast$. It makes $\AAA^{op}$ into a star autonomous category. It is often convenient to include both monoidal structures in the star autonomous signature. 

A star autonomous category is {\em mix autonomous\/} \cite{FleuryR:mix,CockettR:mix} when there is a map $\bot \to \top$. Since there is a natural transformation $X\otimes (Y\parr Z) \tto w (X\otimes Y)\parr Z$ \cite{CockettR:wd}, this induces $X\otimes Z \tto \upsilon X\parr Z$.

An autonomous category is {\em compact\/} when $(X\otimes Y)^\ast  \cong  X^\ast \otimes Y^\ast$ holds naturally in $X,Y$, and $\top^\ast = \top$. 
\end{defn}

\begin{defn}
A {\em \DA\  category\/} is a mix autonomous category also equipped with a\/ {\em lower star\/} functor, i.e. 
\begin{itemize}
\item a covariant strictly monoidal duality $(-)_\ast: \AAA\to \AAA$, such that
\begin{itemize}
\item $\left(X^\ast\right)_\ast =  \left(X_\ast\right)^\ast$ for every object\footnote{Equalities betwen objects are considered {\em "evil"} in categories. We could avoid this by requiring just coherent natural isomorphisms. The next equation between the arrows would then have to be written modulo these isomorphisms.} $X$, and $\top^\ast_\ast = \top$, 
\item $\left(f^\ast\right)_\ast =  \left(f_\ast\right)^\ast$ for every arrow $f$,
\item so that we can write $X^\ddag = X^\ast_\ast$ and $f^\ddag = f^\ast_\ast$;
\end{itemize}
\item a coherent extranatural transformation $X_\ast \tto u X^\ast$, such that the diagram
\[\xymatrix@C-.5pc{
X\otimes X^\ast \ar[dd]_\varepsilon && X\otimes X_\ast \ar[ll]_-{X\otimes u}\ar[rr]^{u'\otimes X} && X^\ddag \otimes X_\ast  \ar[rr]|\sim&& (X^\ast\parr X)^\ddag \ar[dd]^-{\eta^\ddag}
\\ \\
\bot \ar[rrrr] &&&& \top \ar@{=}[rr] && \top^\ddag
}
\]
commutes, where $X\tto{u'} X^\ddag$ is the transpose of $u$.

\end{itemize}
In a\/ {\em dagger compact} category, $u$ is required to be an isomorphism. For simplicity, $X^\ddag = X$ is usually assumed.
\end{defn}

\noindent{\bf Conventions.} Coherence of the above structures means that their natural isomorphisms are unique for the given domain and codomain functors. This means that the structures can be strictified: the natural isomorphisms can be reduced to identities by transferring the functors along them. Assuming that this was done,  $X^{\ast\ast} = X$, $X_{\ast\ast} = X$ will hold on the nose.

\subsection{Dagger compact structure of $\FHilb$}\label{DCS}
We first review the structure of the category of finite-dimensional complex Hilbert spaces and linear maps, since the models in the next sections depend on it, and the needed reconstruction differs from \cite{Abramsky-Coecke:LICS}. 
\begin{itemize}
\item The monoidal structure is given by the usual tensor $\otimes$, with the unit $I=\CCc$.
\item $\HHH^\ast = \FHilb(\HHH, \CCc)$, i.e. the upper star is the dual space functor.
\item $\HHH_\ast$ is the complex conjugate of $\HHH$: it has the same underlying set, but for $z\in \CCc$ and $h\in \HHH$, $z\cdot h$ in $\HHH_\ast$ is $\overline z \cdot h$ in $\HHH$. Any antilinear map from $\HHH$ to $\KKK$ can be viewed as a linear map $\HHH_\ast \to \KKK$ in $\FHilb$.
\item $\HHH^\ddag = \FHilb(\HHH_\ast,\CCc) = \FHilb(\HHH,\CCc)_\ast$. The inner product, viewed as a linear map $<-|->:\HHH_\ast \otimes \HHH \to \CCc$, induces a canonical maps $\HHH\to \HHH^\ddag$ and $\HHH_\ast\to \HHH^\ast$. By the Riesz representation theorem, they are isomorphisms. For $f:\HHH\to \KKK$ and $b\in \KKK$ the image $f^\ddag b\in \HHH$ corresponds along this isomorphism to $\overline{<f^\ddag b|->}\in \FHilb(\HHH,\CCc)_\ast$, defined as the composite $\HHH\tto f \KKK \tto{<b|->} \CCc \tto{\overline{(-)}} \CCc$, i.e.  $<f^\ddag b|-> = <b|f ->$.
\end{itemize}

\subsection{\DDA\  structure of $\int_{\FHilb} \Prop$}
On the objects, the star and dagger functors are
\bear
<\chi\subseteq \HHH>^\ast & = & \left<\chi^\perp \subseteq \HHH^\ast\right>\\
<\chi\subseteq \HHH>_\ast & = & \left<\chi^\perp_\ast \subseteq \HHH_\ast\right>\\
<\chi\subseteq \HHH>^\ddag & = & \left<\chi \subseteq \HHH^\ddag \right>
\eear
where $\chi^\perp = \left\{\psi\in\HHH^\ast\mid \psi\chi = 0\right\}$ is the annihilator, while $\chi^\perp_\ast$ is its inverse image along the Riesz isomorphism $\HHH_\ast\to \HHH^\ast$, obtained by transposing the inner product. The arrow parts of the above functors lift from $\FHilb$, because
\bear
f\chi \subseteq \kappa &\Longrightarrow &f_\ast \chi_\ast \subseteq \kappa_\ast \ \wedge \ f^\ddag \kappa_\ast^\bot \subseteq \chi_\ast^\bot
\eear
The tensors are
\bear
<\chi\subseteq \HHH>\otimes <\kappa \subseteq \KKK> & = & <\chi\otimes \kappa \subseteq \HHH\otimes \KKK>\\
<\chi\subseteq\HHH>\ \parr\  <\kappa\subseteq \KKK> & = &  \left<(\chi^\bot\otimes \kappa^\bot)^\bot\subseteq \HHH\otimes \KKK \right>
\eear
Note that the unit of $\otimes$ is $\top = <\CCc\subseteq \CCc>$, whereas the unit of $\parr$ is $\bot = <0\subseteq \CCc>$. The uniformity is realized by the Riesz' isomorphism.

\section{Comprehending complementarity} \label{Complementarity}
\subsection{Testables}\label{testablesHilb}
A quantum observable is usually represented as a Hermitian operator over a Hilbert space; the actual outcomes of an experiment are the eigenvectors of the Hermitian, which always form a basis. So a simplified view of an observable is that it is a basis of a Hilbert space. The distinctive feature of a quantum observable arises, however, only when we look at several of them at once: it arises from the uncertainty principle, which says that two observables may be {\em incompatible}, or {\em complementary}, in the sense that measuring one disturbs the other. The problem of modeling complementarity is that it is not a one-to-one relationship: many observables may be complementary to many other observables. In order to capture complementarity, we model complementary {\em testables}, construed as families of observables that can be tested together. 

\be{defn}
Let $\HHH$ be a Hilbert space and $\rays{\HHH}$ a set of {\em rays\/} in it, i.e. its 1-dimensional subspaces. The {\em colinearity\/} $\angle (a,b)$ of rays $a,b\in \rays{\HHH}$ is then
\bear
\angle(a,b) & = & \frac{|<x|y>|}{|x||y|}
\eear
for arbitrary nonzero vectors $x\in a$ and $y\in b$. The linearity of the inner product implies that this definition does not depend on the choice of $x$ and $y$. Two rays are orthogonal if their colinearity is 0. They are colinear if it is 1.
\ee{defn}

\be{defn}
For $c\in [0,1]$ and a set $\alpha\subseteq \rays{\HHH}$ of rays in a Hilbert space $\HHH$, the {\em $c$-complement\/} is
\bear
\alpha^c & = & \{x\in \rays{\HHH}\ |\ \forall a\in \alpha.\  \angle(x,a) = c\}
\eear
\ee{defn}

\be{defn} A {\em  testable\/} over a finite-dimensional Hilbert space $\HHH$ is a pair $<\alpha, c>$, where $\alpha\subseteq \rays{\HHH}$, and $c\in [0,1]$, such that
\begin{itemize}
\item both $\alpha$ and $\alpha^c$ span $\HHH$, and
\item $\alpha^{cc} = \alpha$
\end{itemize}
\ee{defn}

\noindent{\bf Examples.} The pair $<\rays{\HHH},0>$ is a testable if and only if $\HHH=0$ is the point, because that is the only case when the $0$-complement $\rays\HHH^0 = \emptyset$ spans $\HHH$. The pair $<\rays\HHH, 1>$ is a testable if and only if $\HHH = \CCc$ is the 1-dimensional space\footnote{The 1-dimensional space is denoted by $\CCc$ because the standard model is usually deployed over complex Hilbert spaces. For our purposes, though, the ground field is largely irrelevant.}, because that is the only case when the $1$-complement $\rays\HHH^1$ is nonempty, and spans $\HHH$. Both $<0,0>$ and $<\CCc,1>$ are self-complementary. For a nontrivial example, let $\HHH$ be an $n$-dimensional Hilbert space and let $c=\frac{1}{\sqrt{n}}$. Then any set $\beta$ of $n$ orthogonal rays $\HHH$ gives a testable $<\beta,c>$, because
\bear
b\in \beta & \iff& \forall x\in \beta^c.\ \angle(b,x) = \frac{1}{\sqrt n}\\
& \iff & \forall x\in \rays{\HHH}. \left(\forall y\in \beta.\ \angle(x,y)  = \frac{1}{\sqrt n}\right)\  \Rightarrow\   \angle(x,b) = \frac{1}{\sqrt n}
\eear 
So if a testable $\beta$ is induced by a basis, then the $c$-complementary testable $\beta^c$ is the union the sets of rays induced by the bases complementary in the sense of \cite{KrausK:uncertainty}.

\medskip

\noindent{\bf Remark.} It is easy to see that $<\alpha^c,c>$ is  testable whenever $<\alpha,c>$ is.

\medskip

\noindent{\bf Why testables?} Since the rays in a testable may not be mutually orthogonal, testing may not allow distinguishing the underlying state \cite[Sec.~2.2.4]{Nielsen-Chuang}. Intuitively, a testable can be tested, but the outcome may not yield a distinct observation. The purpose of testables is to capture complementarity; and  a complement of an observable may be a mixture of multiple observables. The requirement of distinguishability should be imposed later in the development, through the structure of measurements \cite{PavlovicD:QMWS,PavlovicD:Qabs}. 

It is often difficult to provide a clear picture of two ideas in a single structural sweep. Although the complementarity\footnote{The terms "incompatibility", and "unbiasedness" are often used in the same context, sometimes synonymously with complementarity, sometimes in different but related meanings. We take a bird's eye view of complementarity here, and these distinctions do not come about.} of observables and the distinguishability of states are both usually considered in the context of bases, toy models allow us to conceptualize them separately. While distinguishability requires orthogonal families, complementarity does not: a set $\BBB^c$ of all unit vectors complementary to a given basis $\BBB$ is usually not a basis, because some of its elements are not mutually orthogonal. In most models, $\BBB^c$ contain some bases, that can be extracted. But $\BBB$ is in general not completely determined by any of these complementary bases, since each of them usually admits multiple complementary bases. Yet the basis $\BBB$ turns out to be completely determined the complete set $\BBB^c$. That is why it seems more appropriate to model complementarity without the orthogonality requirement, usually imposed on observables.  Hence this attempt with {\em "testables"}.

\smallskip

\noindent{\bf Testables in a category.} Let the specification ${{\HTst}}:\FHilb \to \Rel$ now map each space $\HHH$ to the set ${{\HTst}}\HHH$ of testables over $\HHH$. Each linear map $f\in \FHilb(\HHH,\KKK)$ induces a binary relation $\midd f \subseteq {{\HTst}}\HHH\times {{\HTst}} \KKK$ such that
\bear
<\alpha,c> \midd f <\beta,d> & \iff & f\alpha \subseteq \beta\ \wedge\ f^\ddag \beta^d \subseteq \alpha^c
\eear
The comprehension category $\int_\FHilb {{\HTst}}$ has the triples $<\HHH,\alpha,c>$ as the objects, where $<\alpha,c>$ is a testable over $\HHH$. A morphism $<\HHH,\alpha,c>\to <\KKK,\beta,d>$ is a linear operator $f:\HHH\to \KKK$ which maps  $\alpha$-tests to the $\beta$-tests, while its adjoint $f^\ddag$ maps the complementary $\beta^d$-tests to $\alpha^c$-tests. When the rays in $\alpha$ and $\beta$ are induced by some bases, then this implies that the operator $f$ diagonalizes over these bases. The \DA\  structure is given by
\bear
<\HHH,\alpha,c>^\ast & = & <\HHH, \alpha^c, c>\\
<\HHH,\alpha,c>_\ast & = & <\HHH^\ddag, \alpha^c, c>\\
<\HHH,\alpha,c>^\ddag & = & <\HHH^\ddag, \alpha, c>\\
<\HHH,\alpha,c>\otimes <\KKK,\beta,d> & = & <\HHH\otimes \KKK,\alpha\times \beta, c\cdot d>\\
<\HHH,\alpha,c>\ \parr\  <\KKK,\beta,d> & = & \left<\HHH\otimes \KKK,(\alpha^c\times \beta^d)^{c\cdot d}, c\cdot d\right>
\eear
The unit for both $\otimes$ and $\parr$ is $I = \left<\CCc,\{\CCc\},1\right>$, where $\CCc$ is the tensor unit in $\FHilb$. The mix  map $
<\HHH,\alpha,c>\otimes <\KKK,\beta,d>  \iinclusion\upsilon  <\HHH,\alpha,c>\ \parr\ <\KKK,\beta,d>$
is thus realized by the identity on $\HHH\otimes \KKK$, since $\alpha \times \beta \subseteq (\alpha^c\times \beta^d)^{c\cdot d}$. The uniformity is given by the Riesz map again.

\medskip
\noindent{\bf Entangled vectors live in $\parr$ but not in $\otimes$.} Note that all vectors $I \to <\HHH,\alpha,c>\otimes <\KKK,\beta,d>$ are {\em separated}, i.e. in the form $a\otimes b$, for some $a\in \alpha$ and $b\in \beta$. In contrast, the space $<\HHH,\alpha,c>\ \parr\  <\KKK,\beta,d>$ contains many entangled vectors. In fact, entangled vectors are {\em just\/} those that lie in the complement of the inclusion $\upsilon$ of $\otimes$ into $\parr$. The same phenomenon --- that entanglement can be characterized by the difference between two tensors in a \DA\  category --- was present in {\em general probabilistic theories\/} of Barnum, Barrett, Leifer and Wilce \cite{BarnumH:cloning,BarnumH:teleportation}. Interestingly, this formalism is not based on Hilbert spaces, but on Foulis and Randall's {\em test spaces}. This is what we explore next.

\subsection{Test spaces and testables over relations}\label{tests}
Sets and binary relations form a dagger monidal category $\Rel$, and thus provide a rudimentary model of quantum computation, albeit with mere two scalars. Nevertheless, it turns out that some of the relevant structure in $\Rel$ is rich enough to model some quantum phenomena in a surprisingly informative way. This category is the playground of toy models \cite{SpekkensR:toy,Coecke-Edwards}. Even Foulis' and Randall's {\em test spaces\/} \cite{Foulis-Randall:test,FoulisD:Operational-I,FoulisD:Operational-II}, probably the most extensively studied nonstandard quantum model\footnote{nonstandard in the sense: {\em not\/} the Hilbert space model} \cite{WilceA:testsp}, can be reconstructed and analyzed in this framework, using a comprehension over $\Rel$.

First of all, transferring the algebraic characterization of bases as classical structures from Hilbert spaces \cite{PavlovicD:MSCS08} to relations \cite{PavlovicD:QI09} yields in $\Rel$ a distinction between quantum channels and classical interfaces, which can be used to implement quantum algorithms \cite{PavlovicD:Qabs}\footnote{It should be noted that the exponential speedup of boolean functions, provided by quantum computation, arises from implementing them as unitaries, rather than from some inherent power of relations. But the step of implementing a boolean function as a unitary is just as hard when it is to be executed on a "real" quantum computer, as when it is to be computed as a relation, i.e. executed on a nondeterministic computer.}. The idea is that the  classicality of an abstract basis vector is characterized by its capability to be copied and deleted. In $\Rel$, such a "basis" over a set $X$ corresponds to a partition $X=\coprod_{a\in \alpha} a$, where $\coprod$ denotes the disjoint union, and each $a$ comes with a structure of an abelian group. The abstract basis vectors are the components of the partition, i.e. the disjoint subsets $a \subseteq X$ viewed as the arrows\footnote{In the same way, the vectors in a Hilbert space $\HHH$ are viewed as arrows, i.e. linear operators $I\to \HHH$.} $1\rrel{a} X$ in $\Rel$. See \cite{PavlovicD:QI09} for the details.

Just like the notion of basis, the notion of complementarity can be transferred from $\FHilb$ to $\Rel$. An algebraic characterization of complementary bases in $\FHilb$ was proposed in \cite{Coecke-Duncan}. An equivalent version was transferred to $\Rel$ in \cite{PavlovicD:Qabs}, and complementary bases were used for a relational presentation of a quantum algorithm. A careful exploration of complementary bases in $\Rel$ was provided \cite{PanangadenP:unbiased}. Here we try to comprehend complementarity without the restriction to bases, as explained in the preceding section. So we drop the orthogonality requirement, which in $\Rel$ corresponds to the disjointness of subsets.

\be{defn} A {\em test space\/} over a set $X$ is a family of subsets $\alpha \subseteq \WP X$ which is 
\begin{itemize}
\item covered, i.e. $X = \cup \alpha$, and
\item  irredundant, i.e. $\forall ab\in \alpha.\ a\subseteq b \Rightarrow a=b$
\end{itemize}
The elements of $\alpha$ are called {\em tests}. 
\ee{defn}

\noindent{\bf Remark.} If the tests are disjoint, i.e. $\forall aa'\in \alpha.\  a\cap a' = \emptyset$, then $\alpha$ is a basis in $\Rel$, in the sense of \cite{PavlovicD:QI09}. Tests spaces can thus generalize bases in $\Rel$ in a similar way in which testables generalize observables in $\FHilb$.

\be{defn}
The {\em complement} of a test space $\alpha$ over $X$ is the set of maximal subsets which intersect each test at a single element
\bear
\alpha^\bot & = & \big\{u\in \WP X\ |\ \forall a\in \alpha.\ |u\cap a| = 1\ \wedge\\
&& \hspace{.8em}\forall u'\subseteq u.\ (\forall a\in \alpha.\ |u'\cap a| = 1) \Rightarrow u'= u\big\}
\eear
\ee{defn}

\be{defn}\label{testable}
A family of subsets $\alpha\subseteq \WP X$ is a {\em testable\/} if 
\begin{itemize}
\item both $\alpha$ and $\alpha^\bot$ are test spaces, and
\item $\alpha^{\bot\bot} = \alpha$.
\end{itemize}
\ee{defn}

\noindent{\bf Examples.} The simplest testables over $X$ are provided by the crudest cover $\{X\}$, and by the finest cover $\left\{\{x\}\right\}_{x\in X}$, which happen to be each other's complements. Furthermore, an arbitrary partition $\beta\subseteq \WP X$, i.e. a basis in $\Rel$, is also testable. Its complement $\beta^\bot$ is clearly a test space. It consists of all $p \in \WP X$ which share exactly one element with each $b\in \beta$. Thus all $p\in \beta^\bot$ have the same number of elements, $|p|= |\beta|$. This means that $\beta^\bot$ contains a partition only if $|\beta|$ divides $|X|$. The converse is easily seen to hold. The conclusion is thus that a basis $\beta$ of $X$ in $\Rel$ has a complementary basis if and only if all $b\in \beta$ have the same number of elements, and $|X| = |b|\cdot |\beta|$. Then every complementary basis $\gamma$ has $|\gamma| = |b|$ elements $c\in \gamma$, and each of them has $|c| = |\beta|$ elements. Complementary bases thus form a "rectangular" structure on $X$. This was mentioned and used in \cite[Sec.~5.2]{PavlovicD:Qabs}, and explored in detail in \cite{PanangadenP:unbiased}. But even a basis $\beta$ of $X$ in $\Rel$ that does not admit a complementary basis, {\em viz\/} a complementary observable, always admits a complementary {\em testable}. In terms of the induced equivalence relation 
\bear
x\stackrel\beta\sim y & \iff & \exists b\in \beta.\  x\in b \wedge y\in b
\eear
the requirement that each $p\in \beta^\bot$ must contain exactly one element from each $b\in \beta$ means that $p$ must never contain two $\stackrel\beta\sim$-related elements, and is maximal such, or formally:
\bear
\beta^\bot\ \   = \ \  \big\{p\in \WP X & | & \forall xy\in p.\ x\stackrel\beta\sim y\Rightarrow x=y\ \wedge\\
&& \forall p'\supsetneq p\ \exists xy\in p'.\ x\stackrel\beta\sim y \wedge x\neq y\big\}
\eear
It is easy to see that $\beta^{\bot\bot} = \beta$. In fact, the argument goes through even if $\sim$ is not transitive, i.e. if $\beta$ is not a partition, but the set of maximal cliques for a reflexive symmetric operation. A reader familiar with Girard's {\em coherence spaces\/}  \cite{GirardJY:linear} has by now probably recognized the structure that emerges. Indeed, those testables $\alpha$ that can be presented as sets of cliques of symmetric reflexive relations boil down to coherence spaces\footnote{There, the story is usually told in terms of {\em irreflexive\/} relations. But this is just a matter of convention.}. The \DA\  category of coherence spaces was in fact reconstructed through a comprehension in \cite{PavlovicD:SIC}, and it is now fully embedded in the comprehension category of testables over $\Rel$.

\smallskip
\noindent{\bf Testables over relations.} Let the specification ${{\RTst}}:\Rel \to \Rel$  map each set $X$ to the set ${{\RTst}} X$ of testables over $X$. Each relation $r\in \Rel(X,Y)$ induces a 
relation
\bear
\alpha \midd r \beta & \iff & \WP \WP r(\alpha) \subseteq \beta\ \wedge\ \WP \WP r^{op}(\beta^\bot) \subseteq \alpha^\bot
\eear
where  $\WP r(a) = \{y\in Y\ |\ \exists x\in a.\ xry\}$ and $\WP \WP r(\alpha) =  \{b\in \WP Y\ |\ \exists a\in \alpha.\ \WP r(a) = b\}$ define $\WP\WP X\tto{\wp\wp r} \WP\WP Y$.
The comprehension category $\int_\Rel {{\RTst}}$ has the pairs $<X,\alpha>$ as objects, where $\alpha$ is a testable over $X$. 
A morphism $<X,\alpha>\to <Y,\beta>$ is a relation $r$ which maps every $\alpha$-test to a $\beta$-test, whereas $r^{op}$ maps every $\beta^\bot$-test to an $\alpha^\bot$-test. 

Since $\Rel$ is a compact category with a degenerate dagger structure, the comprehension category $\int_\Rel {{\RTst}}$ is star autonomous category, with a degenerate dagger:
\bear
<X,\alpha>_\ast  & = & <X, \alpha>\\
<X,\alpha>^\ast\ =\ <X,\alpha>^\ddag & = & <X, \alpha^\bot>\\
<X,\alpha>\otimes <Y,\beta> & = & <X\times Y,\alpha\otimes \beta>\\
<X,\alpha>\ \parr\  <Y,\beta> & = & \left<X\times Y,(\alpha^\bot\otimes \beta^\bot)^{\bot}\right>
\eear
where $\alpha\otimes \beta = \{a\times b\ |\ a\in \alpha\wedge b\in \beta\}$. The unit for both tensors is $\left<1,\{1\}\right>$, the only testable over 1. 

\smallskip
\noindent{\bf Remarks.} The resulting category of testables extracts the star autonomous part of the larger category of test spaces and test-preserving relations. This subcategory captures a relational version of complementarity and entanglement, interpreted by analogy with the testables in the preceding section. In contrast with the standard model, we find, e.g., many observables that have complementary testables, but no complementary observables. Is this just an unsound feature of a toy model? Or is the rich, complicated combinatorics of complementary observables in the Hilbert space model just a peculiarity of that model, inessential for quantum computation itself? 

It is fair to also mention that test spaces are interpreted in many different ways in the literature. We viewed them as a relational version of rays\footnote{There is no difference between "rays" and "vectors" in $\Rel$: both are simply subsets.}; many authors view them as an abstraction of bases. This interpretation can also be related using comprehension; but this must be left for another occasion.

\section{Richer comprehensions}\label{More}
\subsection{Quantum probabilities}\label{probabilities}
The comprehension of quantum propositions in Sec.~\ref{propositions} can be generalized to quantum probabilities. While the specification ${\Prop}:\FHilb\to \Rel$ mapped each space $\HHH$ to the lattice ${\Prop} \HHH$ of its closed subspaces, the specification ${\Prob} :\FHilb \to \Rel$ will map each $\HHH$ to the set of quantum probability measures\footnote{This is a simplified version of Mackey's treatment in \cite[Sec.~2.2]{MackeyG:foundations}.} 
\bear
{\Prob} \HHH & = & \{\mu:{\Prop}\HHH\to [0,1]\ |\ \mu(0) = 0\ \wedge\ \mu(\HHH) = \HHH\ \wedge\\
&&\hspace{8.3em}<\chi|\kappa>=0\ \Rightarrow\ \mu(\chi \oplus \kappa) = \mu(\chi)+\mu(\kappa)\}
\eear
where $<\chi|\kappa>=0$ abbreviates $\forall x\in \chi \forall y\in \kappa. <x|y>=0$. The arrow part assigns to every linear map $f\in \FHilb(\HHH,\KKK)$ the relation $\midd f \subseteq {\Prob}\HHH\times {\Prob}\KKK$ which relates the probability measures $\mu$ over $\HHH$ and $\nu$ over $\KKK$ just when it preserves them, i.e.
\bear
\mu \midd f \nu &\iff & \mu = \nu \circ f
\eear
Since a measure on $\HHH$ induces a measure on $\HHH^\ast$, and the measures on $\HHH$ and $\KKK$ induce measures on $\HHH\otimes \KKK$, the \DA\  structure of $\int_\FHilb {\Prob}$ is similar to that of $\int_\FHilb {\Prop}$.

By Gleason's theorem \cite[Sec.~4.2]{Gleason,Piron}, every quantum probability measure (except in dimensions 1 and 2) comes from a density operator, i.e. corresponds to a quantum mixed state. Decomposing density operators as convex combinations of rays, $\int_\FHilb {\Prob}$ can be equivalently presented as the category of mixed states and linear operators that preserve the mixtures. This brings us in the realm of the questions raised in the final sections of \cite{PavlovicD:CQStruct}. It is interesting that the annihilators induce a nontrivial duality on mixed quantum states, displayed in the \DA\  structure of the comprehension category $\int_\FHilb {\Prob}$.

\subsection{Multitestables}\label{multitests}
A test multispace is a test space over a multiset $X$. The idea is that an outcome $x\in X$ may occur several times.
\be{defn}
A\/ {\em test multispace\/} $A$ over $X$ is a pair $A = <\alpha,\omega>$ where 
\begin{itemize}
\item $\alpha \subseteq \WP X$ is a test space, and
\item $\omega : X\to \NNn$ is a function, assigning to each element of $X$ its\/ {\em multiplicity}.
\end{itemize}
We call a test multispace $A = <\alpha,\omega>$ a\/ {\em multitestable} whenever $\alpha$ is a testable.
\ee{defn}

\noindent{\bf Notation.} We define the complement of a test multispace $A=<\alpha,\omega>$ by complementing the underlying test space
\bear
A^\bot & = & \left<\alpha^\bot, \omega\right>
\eear
Conveniently, a test multispace $A$ can also be presented in the "\'etale form" $|A| \tto{A} \WP X$, where
\[
|A| \ \ = \ \ \coprod_{a\in \alpha} \coprod_{x\in a} \omega(x) \ \
 = \ \ \{<a,x,i>\ |\ a\in \alpha\ \wedge\ x\in a\ \wedge\  i\lt \omega (x)\}
\]
and $A$ denotes the projection $<a,x,i>\mapsto a$, by abuse of notation. Note the difference between $|A|$ and $|A^\bot|$.

\smallskip
\noindent{\bf Category of multitestables.} Let the specification ${\MTst}: \Rel \to \Span$ map each set $X$ to the set ${\MTst} X$ of multitestables over it, and each relation $r\in \Rel(X,Y)$ to the span ${\MTst} X \ot \midd r \to {\MTst} Y$, which we view as a matrix of sets, with the entries
\begin{multline*}
A \midd r B\ \ = \ \ \big\{<R,R^\bot>\in \Rel\left(|A|,|B|\right)\times \Rel\left(|B^\bot|,|A^\bot|\right)\ \big|\\
 A; \WP r = R;B\ \wedge\ B^\bot; \WP r^{op} = R^\bot ; A^\bot \big\}
\end{multline*}
\vspace{-1.2\baselineskip}
\[
\xymatrix@R+.3pc@C+.8pc{ |A| \ar[r]^R \ar[d]_A & |B| \ar[d]^B 
& & 
|A^\bot| \ar[d]_{A^\bot} & |B^\bot| \ar[l]_{R^\bot} \ar[d]^{B^\bot}
\\
\wp X \ar[r]_{\wp r} & \wp Y 
& &
\wp X & \wp Y  \ar[l]^{\wp r^{op}}
}
\]

\medskip
The lax structure (\ref{moo}-\ref{eeta}) for $X\rrel r Y$, $Y\rrel s Z$ and $A\in \MTst X$, $B\in \MTst Y$, $C\in \MTst Z$ is given by
\bear
\mu_{rs}^{ABC} : A\midd r B \times B\midd s C & \to &\ \ \ A\midd{r;s} C\\
<R,R^\bot>\ ,\ <S,S^\bot> & \mapsto & \big<R;S\ , \ S^\bot; R^\bot\big>
\eear
whereas $\eta^A\in A\midd{\id_X} A$ is $\left<\id_{|A|}, \id_{|A^\bot|}\right>$.

The objects of the comprehension category $\int_\Rel \MTst$ are the triples $<X, \alpha, \omega>$, $<Y,\beta,\varpi>$, where $<\alpha,\omega>$ is a test multispace over $X$, $<\beta,\varpi>$ over $Y$, etc. A morphism $<X, \alpha, \omega> \to <Y,\beta,\varpi>$ is a triple $\left<r,R,R^\bot\right>$, related as in the above specification. The star autonomous structure, still with the degenerate daggers, is on the objects
\bear
<X,\alpha,\omega>^\ast & = & <X, \alpha^\bot,\omega>\\
<X,\alpha,\omega>\otimes <Y,\beta,\varpi> & = & <X\times Y,\alpha\otimes \beta,\omega\cdot \varpi>\\
<X,\alpha,\omega>\ \parr\  <Y,\beta,\varpi> & = & \left<X\times Y,(\alpha^\bot\otimes \beta^\bot)^{\bot}, \omega\cdot \varpi\right>
\eear
The unit for both tensors is $\left<1,\{1\},1\right>$. The duality on the morphisms is $<r,R,R^\bot>^\ast  =  <r^{op}, R^\bot, R>$ and the monoidal structure is left as an exercise. Nondegenerate dagger structure could be obtained by considering {\em signed\/} multisets, i.e. allowing negative multiplicities, with $\omega:X\to \ZZz$.

\section{Future work}\label{Future}
We presented several categories built by comprehension over $\FHilb$ and $\Rel$, which played the role of the basic models of quantum computation, suitable for refinements and extensions. By refining their dagger compact structure, we arrived in all cases to \DA\  categories. Certain  star autonomous categories have received a lot of attention in research of {\em resource sensitive\/} logics and type systems. Interesting examples of this structure were previously encountered in modeling quantum computation, e.g. in Selinger's explorations of higher order \cite{SelingerP:HOQC}. Abramsky's big toy models \cite{AbramskyS:BigToy,AbramskyS:ToyChu} have star autonomous structure as a prominent feature. 

There is a sense in which dagger autonomous categories give a semantically richer structure than dagger compact categories. They capture not only the abstract composition and duality operations, used for building quantum systems and operations, which correspond to the global operations of the dagger compact structure, but also the complementarity relations, which are modeled in quantum logic, and have been presented in dagger compact categories as local structure. Dagger autonomy arises as soon as complementarity is viewed as a part of the global structure of quantum computation. 

However, a richer or finer picture does not always provide a better insight. One of the most salient features of the dagger compact structure is that the calculations with it are supported by a very convenient string diagram language. Is there a convenient extension of that language catering for the dagger autonomous structure? Its utility may depend on such a language.

But at least some of the fundamental concepts of quantum computation do lift from their simple and robust presentations in dagger compact categories into simple and robust presentations in dagger autonomous categories. For instance, the centerpiece of categorical quantum mechanics \cite{Abramsky-Coecke:LICS}  is the interpretation of entangled pairs in terms of the compact dualities
\[ I\tto{\eta} H^\ast \otimes H \qquad H\otimes H^\ast\tto \varepsilon I\]
which satisfy the adjunction equations
\[(\varepsilon\otimes H)(H\otimes \eta) = \id_H\qquad (H^\ast \otimes \varepsilon)(\eta\otimes H^\ast) = \id_{H^\ast}\]
In star autonomous categories, such dualities lifts to the pairs
\[ \top \tto\eta H^\ast \parr H\qquad H\otimes H^\ast\tto\varepsilon \bot\]
available for every object $H$. The adjunction equations now become
\[
\xymatrix{
H\ar[d]_{H\otimes \eta} \ar[ddrr]^\id & H^\ast \ar[ddrr]_\id \ar[r]^-{\eta\otimes H^\ast}  & (H^\ast \parr H)\otimes H^\ast \ar[dr]^w \\
H\otimes (H^\ast\parr H) \ar[dr]_w &&& H^\ast\parr (H\otimes H^\ast) \ar[d]^{H^\ast \parr \varepsilon}
\\
& (H\otimes H^\ast)\parr H \ar[r]_-{\varepsilon\parr H} & H & H^\ast
}
\]
where $\top \otimes X = X = X\otimes \top$, and $\bot\parr X = X = X\parr \bot$ is assumed for simplicity, and the distributivities $w$ are as in \cite{CockettR:wd,CockettR:mix}. Generalizing the dagger compact view of teleportation \cite{Abramsky-Coecke:LICS,PavlovicD:QMWS,PavlovicD:CQStruct}, one could thus interpret $\top \tto \eta H^\ast \parr H$ as an entangled pair, and $H\otimes H^\ast \tto\varepsilon \bot$ as a basic measurement, and get a rudimentary form of teleportation.  Remarkably, as pointed out at the end of Sec.~\ref{testablesHilb}, in many models entanglement arises {\em exactly\/} from the difference between the two tensors: the entangled pairs live in  $H^\ast \parr H$, the separated pairs in $H^\ast \otimes H$. This phenomenon seems to deserve further thought and exploration.

Proceeding from the above notion of autonomous duality, extending the methods of \cite{PavlovicD:QMWS}, one can show that an associative algebra structure $H\otimes H\tto\nabla H$ in an autonomous category makes $H$ self-dual if it satisfies the autonomous version of the Frobenius condition:
\[\xymatrix{
H\otimes H \ar[d]_-{\Delta \otimes H} \ar[dr]^-{\nabla} \ar[r]^{H\otimes \Delta} & H\otimes(H\parr H) \ar[dr]^-w \\
(H\parr H)\otimes H \ar[dr]_w & H \ar[dr]^\Delta & (H\otimes H)\parr H \ar[d]^-{\nabla \parr H}\\
& H\parr (H\otimes H) \ar[r]_{H\parr \nabla} & H\parr H
}\]
where $\Delta =\nabla^\ddag$. What do classical structures, corresponding to bases and classical observables \cite{PavlovicD:MSCS08,PavlovicD:QI09} lift to in the dagger autonomous framework, and how do their interact with the notions of complement? What is the meaning of the complemenarity of mixed states, touched upon in Sec.~\ref{probabilities}? The comprehension construction provides a handy tool for assembling toy models to explore such questions.

\bigskip

\noindent{\bf Acknowledgement.} Peter Selinger and Chris Heunen provided useful comments and suggestions.

\bibliographystyle{plain}
\bibliography{ref-quantum}

\end{document}